\documentstyle[epsfig]{aipproc}

\begin{document}
\title{The Two-Point Angular Correlation \\ Function and BATSE Sky Exposure}

\author{Xiaoling Chen and Jon Hakkila}
\address{Department of Physics and Astronomy, MSU Box 205 \\
Mankato State University, Mankato, Minnesota 56002}

%\lefthead{LEFT head}
%\righthead{RIGHT head}
\maketitle

\begin{abstract}
The Two-Point Angular Correlation Function is a standard analysis tool
used to study angular anisotropies. Since BATSE's sky exposure (the angular
sampling of gamma-ray bursts) is anisotropic, the TPACF should at some point
identify anisotropies in BATSE burst catalogs due to sky exposure. The effects
of BATSE sky exposure are thus explored here for BATSE 3B and 4B
catalogs. Sky-exposure effects are found to be small.
\end{abstract}

\section*{Introduction}
The Two-Point Angular Correlation Function (TPACF), denoted as
$w(\theta)$, is the frequency distribution of angular separations $\theta$
between celestial objects in the interval $(\theta,\theta+\delta\theta)$. 
The TPACF is normalized to zero when integrated over all separation angles.
A positive value of $w(\theta)$ indicates that objects 
are often found with these
separations, zero indicates no correlation, and a negative value indicates
anti-correlation. The TPACF was developed for studying galaxy clustering
\cite{peebles80}, but has since been applied to gamma-ray burst angular
distributions \cite{hartmann89}. It has been used
extensively in this field, primarily for studying repetition and
clustering \cite{hakkila95,hartmann91,meegan95}. The technique has been
found to be superior in repetition testing to nearest neighbor analysis
\cite{brainerd96}, although it has not been compared directly to detailed
multipole analysis \cite{tegmark96a}. The TPACF has also been used in burst
distance scale studies as an M31 test \cite{hakkila94}. 

Angular sky exposure is important to the modeling of burst angular 
distributions. The gamma-ray burst angular distribution detected by BATSE is 
highly isotropic \cite{meegan96}. However, BATSE's sky exposure is slightly 
anisotropic, and has been specified via a sky exposure map 
\cite{brock92,hakkila98}.
This anisotropy is due primarily to earth blockage. It causes a small 
systematic deviation from isotropy that could show up in the gamma-ray burst 
angular distribution.

In this study, we identify the form of the TPACF taken by BATSE 
sky exposure in the case of a large (infinite) number of detected bursts.
We also search the BATSE 3B and 4B catalogs to determine to what extent this
effect is present.

\section*{Analysis Method}

Two methods have been used to calculate the TPACF while including sky
exposure effects: (a) Monte Carlo simulations of random burst catalogs 
are performed using sky exposure as a mask, and (b) summation of isotropic 
sampled points with the declination-dependent sky exposure are performed.

The first technique requires simulation of gamma-ray burst catalogs,
using an assumed isotropic gamma-ray burst angular distribution.
The sky exposure function is used as 
a mask to selectively decide whether or not a simulated burst was 
detected. Using the detected bursts, the number of distinct burst 
pairs with different angular separations $(\theta,\theta+\delta\theta)$ is
identified, and is used to calculate the TPACF via:

\begin{equation}
1+w(\theta)=\frac{2 n_p \Omega}{N(N-1)\langle\delta\Omega\rangle}
\end{equation}  

where $n_p$ is the expected number of pairs with separations 
$\theta$ to $\theta+\delta\theta$, and $\langle\delta\Omega\rangle$ is 
the mean value of the solid angle subtended by the annulus $\theta$ to 
$\theta+\delta\theta$.
This form of the TPACF $w(\theta)$ is valid when it is
estimated from a discrete set of sky objects \cite{peebles80}, that is,
via a limited number of detected bursts. Statistical variations between
Monte Carlo catalogs are removed by numerous Monte Carlo runs using the
same number of detected bursts.

The second technique calculates $w(\theta)$ for a large number of 
isotropically-sampled points using a weighted sky exposure. Equal sampling
is obtained by using a technique for pixelizing the celestial sphere
in a smooth and regular way \cite{tegmark96b}.The total number of 
observed bursts $N$ can be described in terms of the sky exposure $e_i$
at each sampled sky location $i$ as
\begin{equation}
N=A\sum_{i=1}^Ne_{i}
\end{equation}
where $A$ is a normalization constant. Notice that when all sky regions
are completely-sampled ($e_{i}=1$ everywhere), then $A=1$. Otherwise, 
$A>1$. For each burst detected, 
$\frac{1}{e_{i}}$ bursts should have been detected.
Similarly, for each evenly-sampled point on the celestial sphere, 
bursts should be observed only a fraction of the time $e_{i}$. 
The weighted exposure $y_{i}$ is
\begin{equation}
y_{i}=\frac{N e_{i}}{\sum_{i=1}^Ne_{i}}=A e_{i},
\end{equation}
and
\begin{equation}
\sum_{i=1}^Ny_{i}=N
\end{equation} 

Equation (1) is used to calculate the TPACF $w(\theta)$ for
discrete points, with the number of pairs now given as
\begin{equation}
n_{p}(\theta)=\sum_{i=1}^N\sum_{j=i+1}^{N}y_{i}y_{j}=A^2\sum_{i=1}^Ne_{i}\sum_{j=i+1}^{N}e_{j}
\end{equation}
This technique produces an exact solution in the limit $N\rightarrow\infty$.
% \cite{hartmann91}:
%\begin{equation}
%1+w_{ij}(\theta)=\int{d\vec{\Omega}_i}\int{d\vec{\Omega}_j 
%p(\vec{\Omega}_i) p(\vec{\Omega}_j) \delta(\vec{\Omega}_i\vec{\Omega}_j
%-|\vec{\Omega}_i||\vec{\Omega}_j|\cos\theta)}
%\end{equation}
%where $\vec{\Omega}_i$ and $\vec{\Omega}_j$ are the vectorial solid angles 
%of the two regions $i$ and $j$, $p(\vec{\Omega}_i)$ and $p(\vec{\Omega}_j)$
%are the detection probabilities of the $i$th and $j$th bursts, and sums are
%taken over the $N$ observed points. The probabilities $p(\vec{\Omega}_i)$ 
%and $p(\vec{\Omega}_j)$ correspond to the normalized sky exposures of
%the $i$th and $j$th points.

A total of 6252 sampling points is used for the equal sampling technique,
with the separation between any two neighboring
points being $1.45^\circ$. This separation is smaller than the $1.5^\circ$
BATSE systematic burst localization error and has been chosen because
localization uncertainties make it difficult to obtain 
meaningful data from smaller angular scales.

Some systematic errors in the TPACF are still found using the equal 
sampling technique. These are due to the discrete placement of pixel 
separations into $w(\theta)$ angular bins. Exact corrections can
be obtained for the values in each angular bin, since for ideal exposure 
$e_i = 1$ the number of expected pairs per bin can be
found from direct integration.

\section*{Results and Discussion}

\begin{figure}[b!] % fig 1
\centerline{\epsfig{file=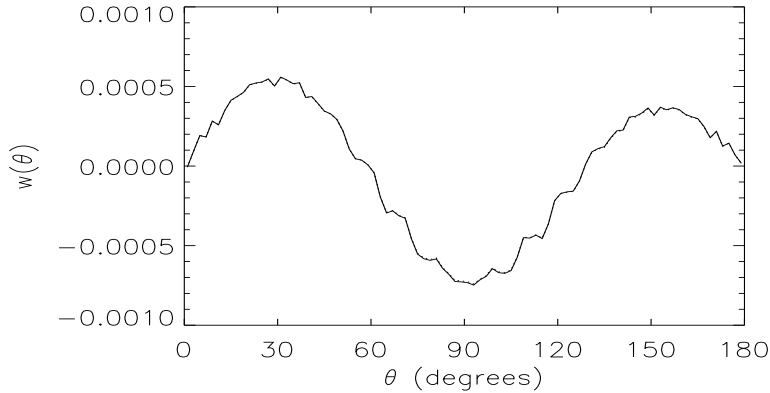,height=2.0in,width=3.5in}}
\vspace{12pt}
\caption{Two-Point Angular Correlation Function indicating 
the extent of BATSE 3B (dotted line) and 4B (solid line) sky exposure
using the equal sampling method.}
\label{fig1}
\end{figure}

Figure \ref{fig1} demonstrates application of the equal sampling technique 
to the BATSE 3B and 4B catalogs. Differences between the 3B and 4B exposures
are insignificant. The Monte Carlo technique produces similar 
results when averaged over a large number of Monte Carlo runs, although
computational constraints limited us to ten runs per model.

The effects of sky exposure in the TPACF are found to be very small,
although they produce a distinct TPACF functional form. Figure \ref{fig1} 
exhibits a depletion of burst pairs for $60^\circ \leq \theta \leq 120^\circ$,
with corresponding pair excesses for $\theta < 60^\circ$ and
$\theta > 120^\circ$. The pair excesses disappear in the
limits of small ($\theta \rightarrow 0^\circ$) and large ($\theta \rightarrow 
180^\circ$) separation angles. There is a slight enhancement in the
peak at $\theta \approx 30^\circ$ relative to that at $\theta \approx
90^\circ$.

Enhanced polar to equatorial coverage produces a lower angular density band 
around the celestial equator along with two higher angular-density polar 
regions. Bursts form an excess number of pairs on
angular scales $\theta \approx 30^\circ$ and $\theta \approx 150^\circ$
since both polar and equatorial bursts sample a larger number of their
pair partners from the enhanced polar regions on these angular scales. In
contrast, the total number of burst pairs near $\theta \approx 90^\circ$ is
depleted because polar region bursts have relatively few neighbors on
angular scales of $\theta \approx 90^\circ$ while equatorial bursts produce 
some $\theta \approx 90^\circ$ pairs from polar bursts along with others from
the lower-density equatorial region. On small angular scales ($\theta \approx
0^\circ$) bursts draw their companions from the surrounding region with a
similar angular density; thus the excess of polar burst pairs is balanced
overall by the depletion of equatorial burst pairs. The same reasoning
applies to large angular scales ($\theta \approx 180^\circ$), since bursts
also draw their companions from a region of similar density on this angular
scale. The enhanced peak at $\theta \approx 30^\circ$ (relative to that at 
$\theta \approx 90^\circ$) is probably due to better northern hemisphere 
exposure.

When properly normalized, the distribution of distinct pairs found 
in all angular annuli $(\theta,\theta+\delta\theta)$ becomes the
TPACF. The number of distinct pairs in each annulus can be used to
identify the importance of sky exposure effects on the BATSE 
catalogs, since the number of bursts is needed to obtain the 
measurement error. A $\chi^2$ value can be found by comparing
the BATSE distinct pair distribution with the number expected
(a) in the absence of sky exposure, and (b) in the presence
of sky exposure.

The results of this analysis are obtained using the two methods
described previously, and are demonstrated in Table \ref{table}.
The reduced $\chi^2_\nu$ values are given for
the BATSE 3B and 4B catalogs both when sky exposure
effects are not taken into account (e.g. an
isotropic distribution) and when these effects are included.
Different data binnings have also been used in order to
determine their effect on $\chi^2_\nu$. The results do
not depend strongly on data binning, whether based on
the sizes or on angular offsets of the bins.

\begin{table}
\caption{Effects of sky exposure on the BATSE 3B and 4B catalogs. The reduced 
$\chi^2_\nu$ values shown are obtained by comparing the TPACF ($\nu = 89$
degrees of freedom) for the specified catalog to models accounting for 
exposure and those not accounting for it ({\em e.g.} $w(\theta) = 0$ 
everywhere). The calculations are performed using both Monte Carlo analysis and
the equal sampling technique.
}

\label{table}
\begin{tabular}{||c||c|c|c|c||}
Technique & 3B no exposure & 3B exposure & 4B no exposure & 4B
exposure \\
\tableline
Monte Carlo & 1.12 & 1.03 & 1.47 & 1.30 \\
equal sampling & 0.94 & 1.01 & 1.12 & 1.08 \\
\end{tabular}
\end{table}

No significant improvement in $\chi^2_\nu$ is obtained when sky exposure
is taken into account, as indicated by $\chi^2_\nu \approx 1$ for all models.
This result is not terribly surprising since statistical errors
due to burst counting statistics produce errors that are typically 
$10$ to $100$ times as large as the maximum TPACF amplitude of the 
sky exposure (for data binned on $2^\circ$ intervals).

However, other anisotropy tests appear to be capable of detecting
sky exposure. For example, dipole and quadrupole
moments of exposure are marginally detectable \cite{meegan96}.
There are several possible explanations for the apparent inadequacy of
the TPACF to statistically detect exposure effects:
\begin{itemize}
\item Anisotropies are more statistically-significant when described 
using the coordinate system for which the anisotropy is largest. Generalized 
coordinates (such as $\theta$) lose some of this significance 
by not taking advantage of this {\em a priori} knowledge.
\item By using only the two-point angular correlation function, rather 
than all n-point orthogonal functions, we are not making use of all 
available information. Just as the dipole moment can identify some anisotropies
to which the quadrupole moment is insensitive, it may be that the TPACF
is less sensitive to sky exposure effects than the higher-order
correlation functions. Limiting the analysis to just one component of
all correlation functions may also limit the ability of the test to identify
BATSE sky exposure in the gamma-ray burst angular distribution.
\end{itemize}

\end{document}